\documentclass[12pt]{article}
\usepackage{amssymb,amsmath,epsfig}
\allowdisplaybreaks

\begin{document}

\title{\bf Viscous Chaplygin
Gas Models as a Spherical Top-Hat Collapsing Fluids}
\author{Abdul Jawad$^1$ \thanks{jawadab181@yahoo.com;~~abduljawad@ciitlahore.edu.pk} and
Ayesha Iqbal$^{1,2}$
\thanks{ayeshausmann@yahoo.com}\\
$^1$ Department of Mathematics, COMSATS Institute of\\ Information
Technology, Lahore-54000, Pakistan.\\
$^{2}$ Department of Mathematics, Govt. College University,\\
Faisalabad, Pakistan.}

\date{}

\maketitle
\begin{abstract}
We study the spherical top-hat collapse in Einstein gravity and loop
quantum cosmology by taking the non-linear evolution of viscous
modified variable chaplygin gas and viscous generalized cosmic
chaplygin gas. We calculate the equation of state parameter, square
speed of sound, perturbed equation of state parameter, perturbed
square speed of sound, density contrast and divergence of peculiar
velocity in perturbed region and discussed their behavior. It is
observed that both chaplygin gas models support the spherical
collapse in Einstein as well as loop quantum cosmology because
density contrast remains positive in both cases and the perturbed
equation of state parameter remains positive at the present epoch as
well as near future. It is remarked here that these parameters
provide the consistence results for both chaplygin gas models in
both gravities.
\end{abstract}
Loop quantum cosmology; Density contrast.\\

\section{Introduction}

The discovery of accelerating expansion of the universe is a
milestone for cosmology. This acceleration is generally believed to
be caused by the vacuum energy or exotic matter called "dark energy"
(DE) \cite{1,2}. This exotic matter has positive energy density and
strong negative pressure which can be represented by an equation of
state (EoS) parameter $\omega =\frac{p}{\rho}<\frac{-1}{3}$. The
simplest candidate of DE is cosmological constant ($\Lambda$) having
EoS parameter $\omega=-1$ \cite{3}. Other candidates are
quintessence ($-1<\omega<\frac{-1}{3}$) \cite{4} and phantom
($\omega<-1 $) \cite{5}. In the universe, the contribution of DE in
energy density is about 73\% and an unknown form of matter called
"dark matter" (DM) is about 23\%. The remaining 4\% of the energy
density corresponds to ordinary known matter. DM (in its cold
version) is a dust-like fluid with no pressure. It is attractive in
nature and can not be seen by a telescope. It do not absorb or emit
light or any gravitational waves. The existence of this type of
matter has been proved by gravitational effect on visible matter and
gravitational lensing of background radiation.

The Chaplygin gas (CG) has been presented which possesses the
unified picture of DE and DM. It was introduced by Chaplygin who
studied it in hydrodynamical context \cite{6}. This gas was not
consistent with angular power spectrum of the cosmic microwave
background (CMB) or the angular distance scale of the baryon
acoustic oscillations (BAO) \cite{7,8}. Therefore, an extension of
CG model named as generalized CG (GCG) has been proposed. This model
corresponds to almost dust $(p=0)$ at high density which does not
agree completely with the universe. Then GCG is extended to modified
CG (MGCG). The MGCG is more appropriate choice to have constant
negative pressure at low energy density and high pressure at high
density. The viscous CG model is the suitable candidate of unified
DE and cold DM as a unique imperfect fluid \cite{9,10}. It has bulk
viscosity with negative pressure suggested by observations
\cite{9,10}. The Generalized Cosmic Chaplygin Gas (GCCG) is another
form of DE having the consequence of an accelerating phase of
universe, which was introduced by Gonzalez-Diaz \cite{18}. This
model is stable and free from unphysical behaviour even when the
vacuum fluid satisfies the phantom energy condition.

The ``cosmological collapse", a mechanism proposed by physicists
predicts that the universe will soon stop expanding and collapse in
itself. The spherical collapse (SC), introduced by Gunn and Gott
\cite{11} provides a way to glimpse into the nonlinear regime of
perturbation theory. It explains how a small spherical patch of
homogeneous overdensity forms a bound system via gravitation
instability \cite{12}. It describes the evolution of a spherically
symmetric perturbation embedded in a static, expanding or collapsing
homogeneous background.
One assumes a spherical 'top hat' profile for the
perturbed region, i.e. a spherically symmetric perturbation in some
region of space with constant density \cite{top1}. The assumption of
a 'top hat' profile leads to the SC model as the uniformity of the
perturbation is maintained throughout the collapse, making its
evolution only time dependent. As a consequence, we do not need to
worry about gradients inside the perturbed region. Thus, the STHC
identifies the evolution of a homogeneous mini-universe inside a
larger homogeneous universe.

Specifically, it is suggested that the non-linear effects of GCG can
not be ignored because it can produce a back-reaction in the
background dynamics, leads to crucial constraints on the validity of
linear theory as soon as the first scales become non-linear. In
\cite{top2}, the authors have studied the non-linear evolution of
dark matter and dark energy by taking CG model, using
generalizations of the spherical model that incorporated effects of
the acoustic horizon. An interesting phenomenon was found there: a
fraction of the CG condensated and never reached a stage where its
properties changed from dark-matter-like to dark-energy-like. A
fully non-linear analysis is a cumbersome task usually handled by
hydrodynamical/Nbody numerical codes (see e.g. \cite{top1,top3}).

Moreover, some works have been done by assuming spherically
symmetric perturbation in some region of space with constant density
as a spherical top-hat profile for the perturbed region. Fabris et
al. \cite{14} studied the evolution of density perturbation in a
universe dominated by CG. Their model gave the required density
contrast observed in large scale structures of universe in Newtonian
approach. Carturan and Finelli \cite{15} conducted same
investigations for GCG. Hence, Fernandes et al. \cite{top1} studied
the evolution of density perturbation in a universe dominated by CG.
Their model gave the required density contrast observed in large
scale structures of universe in Newtonian approach. Carturan and
Finelli \cite{15} conducted same investigations for GCG. Crames et
al. \cite{17} investigated the STHC model using viscous GCG (VGCG).
Li and Xu \cite{9,10} have extended to above work by adding bulk
viscosity in GCG model. They analyzed effect of bulk viscosity on
structure formations of the GCG model having spherically symmetric
perturbations.

Recently, Karbasi and Razmi \cite{17aa} have discussed the STHC in
the presence of MCG. Also, Ujjal and Mubasher \cite{13} have
analyzed the STHC scenario in the presence of viscus MCG in Einstein
and loop quantum gravities. Motivated by these works, we study the
STHC in the presence of viscous modified variable CG(VMVCG) and
viscous GCCG (VGCCG) in Einstein as well as loop quantum gravities.
We organize our paper as follows: In section \textbf{2}, we give a
discussion of VMVCG and corresponding STHC scenario in both
gravities. Section \textbf{3} provides STHC of VGCCG. We conclude
the results in Section \textbf{4}.

\section{Viscous Modified Variable Chaplygin Model}

We consider the flat FRW universe as
\begin{eqnarray}\label{1}
ds^{2}=-dt^{2}+a^{2}(t)[dr^{2}+r^{2}d\theta^{2}+r^{2}\sin^{2}\theta
d\phi^{2}],
\end{eqnarray}
where $a(t)$ is the scale factor of the universe. The Einstein field
equations are given by
\begin{eqnarray}\label{2}
H^{2}=\frac{8\pi G}{3}\rho,\quad \dot{H}=-4\pi G(\rho+p),
\end{eqnarray}
Next, we take $8\pi G = 1$. The CG acts as a pressureless fluid for
small values of the scale factor and tends to accelerated expansion
for large values of the scale factor. The EoS for this model has
following form
\begin{equation}\nonumber
p_{d}=-\frac{B}{\rho},
\end{equation}
where $B$ is a positive constant. The EoS parameter for CG has been
generalized to the form \cite{19}-\cite{21}.
\begin{equation}\nonumber
p_{d}=-\frac{B}{\rho^{\alpha}},
\end{equation}
with $ 0<\alpha\leq1$. The EoS of CG has an interesting connection
with the D-branes which are expressed via Nambu-Goto action
\cite{14}. It also enjoys connections with the Newtonian
hydrodynamical equations. Further the Eddington-Born-Infeld model
can be seen as an affine connection version for the CG approach
\cite{22}. The CG has been extensively studied within the unified
DE-DM models. The CG was extended to MVCG with following EoS
\begin{equation}\label{3a}
p_{d}=\tilde{A}\rho-\frac{B(a)}{\rho^{\alpha}},
\end{equation}
where $B(a)=B_{0}a^{-m}$  with $m$ a non-negative integer and
$\tilde{A}$ is a constant constrained by the astrophysical data. The
special case $\tilde{A}=\frac{1}{3}$ is the best fitted value to
describe evolution of the universe from radiation regime to
$\Lambda$-cold DM regime.

The MVCG satisfactorily accommodates an accelerating phase as well
as matter dominated phase of the universe. It is also consistent
with the observational studies dealing with the large scale
structure \cite{23}. We assume here that the spacetime is filled
with only one component fluid having a bulk viscosity. This
component is defined in terms of effective pressure $p$ as follows
\begin{equation}\label{3}
p=p_{d}+\Pi,
\end{equation}
which is the sum of the equilibrium pressure $p_{d}$ and the bulk
pressure $\Pi=-\xi_{;\gamma}^{\gamma}$ where $u^{\gamma}$ is the
four velocity of the fluid and $\xi$ represents the coefficient of
bulk viscosity (which is a function of energy density). The first
attempts at creating a viscosity theory of relativistic fluids were
executed by Eckart \cite{24a} and Landau and Lifshitz \cite{25bb}.
They considered only a first-order deviation from equilibrium. The
bulk viscous pressure $p_{d}$ is represented by Eckart's expression
which is proportional to the Hubble parameter $H$. The
proportionality factor identified as the bulk viscosity coefficient
$\xi=\xi_{0}\rho^{e}$ where $\xi_{0}$ and $e$ are constants. For
simplicity, choosing $e=\frac{1}{2},~\Pi$ can be written as
\cite{7},
\begin{equation}\label{3b}
\nonumber\Pi=-3\xi_{0}H\sqrt{\rho},
\end{equation}
where $H=\frac{\dot{a}}{a}$ is the Hubble parameter with dot
representing time derivative.

By combining Eqs.(\ref{3a})-(\ref{3b}), we obtain VMVCG
\begin{equation}\label{5}
p=\tilde{A}\rho-\frac{B_{0}a^{-m}}{\rho^{\alpha}}-3\xi_{0}H\sqrt{\rho}.
\end{equation}
The conservation equation is given by
\begin{equation}\label{6}
\dot{\rho}+3H(\rho+p)=0.
\end{equation}
By inserting Eq.(\ref{5}) in (\ref{6}), we get the following
solution
\begin{eqnarray}\label{7}
\rho=\bigg(\frac{3B_0(1+\alpha)}{3(1+\alpha)(1+\tilde{A}-\sqrt{3}\xi
_0)-m}a^{m}+\frac{C}{a^{3(1+\alpha)(1+\tilde{A}-\sqrt{3}\xi
_0)}}\bigg)^\frac{1}{\alpha+1}.
\end{eqnarray}
In terms of redshift, i.e. using $a=\frac{1}{1+z}$, we obtain
\begin{eqnarray}\label{7}
\rho(z)&=&\rho_0\bigg(\frac{3B_0(1+\alpha)(1+z)^{m}}{3(1+\alpha)(1+\tilde{A}-\sqrt{3}\xi
_0)-m}+C(1+z)^{3(1+\alpha)(1+\tilde{A}-\sqrt{3}\xi
_0)}\bigg)^\frac{1}{\alpha+1}.
\end{eqnarray}
By inserting the above expression in field equation, we will get the
Hubble parameter as follows
\begin{eqnarray}\nonumber
H(z)&=&H_0\bigg[\bigg(\frac{3B_0(1+\alpha)(1+z)^{m}}{3(\alpha+1)(1+\tilde{A}-\sqrt{3}\xi_0)-m}
C(1+z)^{3(\alpha+1)(1+\tilde{A}-\sqrt{3}\xi_0)}\bigg)^{\frac{1}{\alpha+1}}\\\label{10}&\times&\Omega_0
\bigg]^{\frac{1}{2}}.
\end{eqnarray}
The EoS parameter becomes
\begin{equation}\label{8}
\omega=\frac{p}{\rho}=\tilde{A}-\frac{B_{0}a^{-m}}{\rho^{\alpha+1}}-\frac{3\xi_{0}H}{\sqrt{\rho}}.
\end{equation}
The adiabatic sound speed can be defined as follows
\begin{equation}\label{9}
c_{s}^{2}=\frac{dp}{d\rho}.
\end{equation}
For VMVCG, this parameter can be obtained by using Eqs.
(\ref{5})-(\ref{10}).

\subsection{STHC for VMVCG in Einstein Gravity}

Following the assumption of a top-hat profile, the density
perturbation is uniform throughout the collapse. In this case the
evolution of perturbation is only time-dependent. In STHC model, the
background evolution equation are in following form
\begin{eqnarray}\nonumber
\dot{\rho}=-3H(\rho+p), \frac{\ddot{a}}{a}=\frac{-4 \pi
G}{3}\sum_{i}(\rho_{i}+ p_{i}).
\end{eqnarray}
The perturbed quantities $\rho_{c}$ and $p_{c}$ are related to their
background counterparts by $\rho_{c}=\rho+\delta\rho$ and
$p_{c}=p+\delta p$. Applying the perturbations, the EoS for VMVCG
becomes
\begin{equation}\label{15}
\omega_{c}=\frac{\tilde{A}-\frac{B_{0}a^{-m}}{\rho^{\alpha+1}}-\frac{3\xi_{0}H}{\sqrt{\rho}}+\delta
c_{e}^{2}}{1+\delta}.
\end{equation}
The perturbed equation of sound speed leads to
\begin{equation}\nonumber
c_{e}^{2}=\frac{p_{c}-\rho\tilde{A}+\frac{B_{0}
a^{-m}}{\rho^{\alpha}}+3\xi_{0}H \sqrt{\rho}}{\rho_{c}-\rho},
\end{equation}
which implies
\begin{equation}\label{16}
c_{e}^{2}=\acute{A}-\frac{B_{0}(1+z)^{-m}(1-(1+\delta)^{\alpha})}{\delta
\rho^{(1+\alpha)}
(1+\delta)^{\alpha}}-\frac{3}{\delta\sqrt{\rho}}H\xi_{0}(\sqrt{1+\delta}-1).
\end{equation}
The basic equations about the density contrast and divergence of
peculiar velocity in perturbed region are presented in the
\textbf{Appendix}. Hence, by following the procedure of
\cite{17aa,13}, we can get density contrast and divergence of
peculiar velocity for VMVCG in the following for Einstein gravity
\begin{eqnarray}\nonumber
\frac{d\delta}{dz}&=&
\frac{3}{1+z}\bigg(\tilde{A}-\frac{B_{0}(1+z)^{-m}(1-(1+\delta)^{\alpha})}{\delta
\rho^{1+\alpha}(1+\delta)^{\alpha}}-\frac{3}{\delta\sqrt{\rho}}H\xi_{0}(\sqrt{1+\delta}-1)\\\nonumber
&-&\omega\bigg)\delta
-\bigg(1+\omega+\bigg(1+(\tilde{A}-\frac{B_{0}(1+z)^{-m}(1-(1+\delta)^{\alpha})}{\delta
\rho^{1+\alpha}
(1+\delta)^{\alpha}}-\frac{3}{\delta\sqrt{\rho}}H\xi_{0}\\\label{22}
&\times&(\sqrt{1+\delta}-1))\bigg)\delta\bigg)\frac{\theta}{H},
\\\nonumber
\frac{d\theta}{dz}&=& \frac{\theta }{1+z}+\frac{\theta^{2}}{3
H(z)}+\frac{3H(z)}{2(1+z)^{2}}\bigg(1+3
\bigg(\tilde{A}-\frac{B_{0}(1+z)^{-m}\big(1-(1+\delta)^{\alpha}\big)}{\delta
\rho^{1+\alpha}
(1+\delta)^{\alpha}}\\\label{23}&-&\frac{3}{\delta\sqrt{\rho}}H\xi_{0}(\sqrt{1+\delta}-1)\bigg)\bigg)\delta
\Omega.
\end{eqnarray}

\subsection{STHC for VMVCG in Loop Quantum Cosmology}

The LQC is a quantization of symmetry reduced spactetime. Some
phenomenon like predictions of cosmic inflation in the early
universe \cite{36}, late time cosmic acceleration \cite{37} and
primordial gravitational waves \cite{38} have also explored in this
gravity. In this gravity, the cosmological perturbation theory has
also investigated \cite{39}. The LQC possesses the properties of
non-perturbative and background independent quantization of gravity
\cite{L22}-\cite{L27}. Various DE models have been investigated in
this gravity \cite{L27a,L27b}. Jamil et al. \cite{L29} have
investigated the cosmic coincidence problem by assuming the MCG
coupled to DM. It has also been found that the future singularity
appearing in the standard FRW cosmology can be avoided by loop
quantum effects \cite{L30}. Chakraborty et al. \cite{L31} have
tested MCG in LQC.

The modified Einstein's field equations in LQC for FRW metric are
given by
\begin{eqnarray}\label{24}
H^{2}&=& \frac{8\pi G \rho}{3}\left(1-\frac{\rho}{\rho_{1}}\right),
\\\label{25} \dot{H}&=& -4\pi G (\rho +
p)\left(1-\frac{2\rho}{\rho_{1}}\right),
\end{eqnarray}
where $\rho_{1}= \frac{\sqrt{3}}{16\pi^{2}\gamma^{3} G^{2} h}$ is
called the critical loop quantum density, $\gamma$ is the
dimensionless Barbero-Immirzi parameter.

In this case, the Hubble parameter is obtained as
\begin{eqnarray}\nonumber
H(z)&=&H_{0}\left[\Omega_{0}\frac{3B_{0}(1+\alpha)(1+z)^{m}}{3(1+\alpha)(1+\tilde{A}-3\xi_{0})
-m}+c(1+z)^{3(1+\alpha)(1+\tilde{A}-3\xi_{0})}\right]^{\frac{1}{1+\alpha}}\\\nonumber
&\times&\left[1-\frac{\rho_{0}}{\rho_{1}}\left(\frac{3B_{0}(1+\alpha)(1+z)^{m}}{3(1+\alpha)
(1+\tilde{A}-3\xi_{0})-m}
+c(1+z)^{3(1+\alpha)}(1+\tilde{A}\right.\right.\\\label{26}&-&\left.\left.3\xi_{0})^{\frac{1}{1+\alpha}}\right)
\right]^{\frac{1}{2}}.
\end{eqnarray}
In this case, the EoS parameter (10) becomes
\begin{equation}\label{27}
\omega=\tilde{A}-\frac{B_{0}a^{-m}}{\rho^{\alpha+1}}-\frac{3\xi_{0}H_{0}
\sqrt{\frac{\Omega_{0}\rho}{\rho_{0}}(1-\frac{\rho}{\rho_{1}}})}{\sqrt{\rho}}.
\end{equation}
The adiabatic sound speed becomes
\begin{eqnarray}\label{28}
c_{s}^{2}=\tilde{A}-\frac{\alpha
B_{0}a^{-m}}{\rho^{\alpha+1}}-\frac{3\xi_{0}}{2
\sqrt{\rho}}{H_{0}\sqrt{\frac{\Omega_{0}\rho}{\rho_{0}}(1-\frac{\rho}{\rho_{1}})}},
\end{eqnarray}
the perturbed EoS (14) turns out to be
\begin{equation}\label{29}
\omega_{c}=\frac{1}{1+\delta}\bigg(\tilde{A}-\frac{B_{0}a^{-m}}{\rho^{\alpha+1}}-\frac{3\xi_{0}H}{\sqrt{\rho}}+\delta
c_{e}^{2}\bigg).
\end{equation}
The perturbed equation of sound speed in terms of redshift takes the
form
\begin{eqnarray}\label{32}
c_{e}^{2}&=&\acute{A}-\frac{B_{0}(1+z)^{-m}(1-(1+\delta)^{\alpha})}{\delta
\rho^(1+\alpha)
(1+\delta)^{\alpha}}-\frac{3\xi_{0}}{\delta\sqrt{\rho}}H_{0}\sqrt{\frac{\Omega_{0}\rho}{\rho_{0}}(1-\frac{\rho}{\rho_{1}})}
\\\nonumber&\times&(\sqrt{1+\delta}-1).
\end{eqnarray}

The dynamical equations of density contrast $\delta$ remains the
same whereas equation for $\theta$ for MVGCG in LQC becomes
\begin{eqnarray}\nonumber
\frac{d\theta}{dz}&=& \frac{\theta }{1+z}+\frac{\theta^{2}}{3
H(z)}+\frac{3H(z)}{2(1+z)^{2}}\bigg(1+3
\bigg(\tilde{A}-\frac{B_{0}(1+z)^{-m}(1-(1+\delta)^{\alpha})}{\delta
\rho^(1+\alpha) (1+\delta)^{\alpha}}\\\nonumber
&-&\frac{3}{\delta\sqrt{\rho}}H\xi_{0}(\sqrt{1+\delta}-1)\bigg)\bigg)\delta
\Omega\bigg((1+3 c_{e}^{2})-\frac{6\Omega H^{2}}{
\rho_{1}}\bigg(3(1+\delta)(1+3c_{e}^{2})\\\label{34}
&+&(3\omega-\delta+1)\bigg)\bigg).
\end{eqnarray}
\begin{figure}[h]
\begin{minipage}{14pc}
\includegraphics[width=16pc]{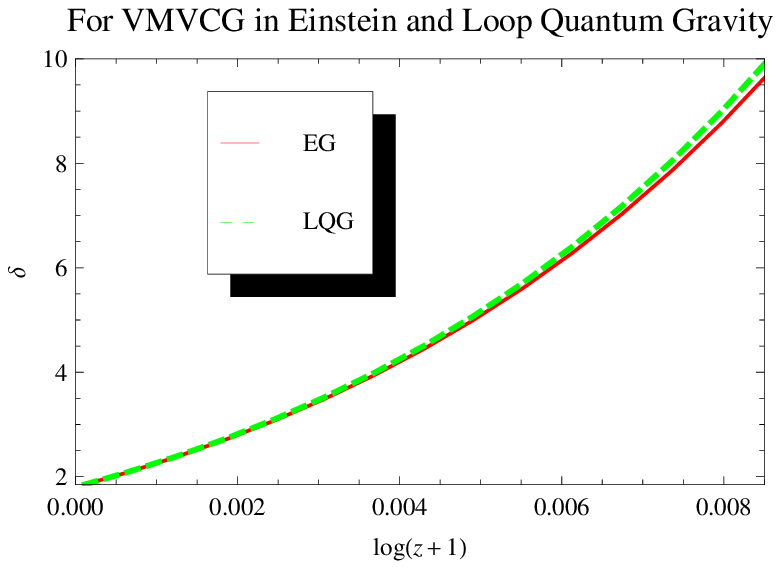}
\caption{\label{label}Plot of $\delta$ versus $\log(1+z)$.}
\end{minipage}\hspace{3pc}%
\begin{minipage}{14pc}
\includegraphics[width=16pc]{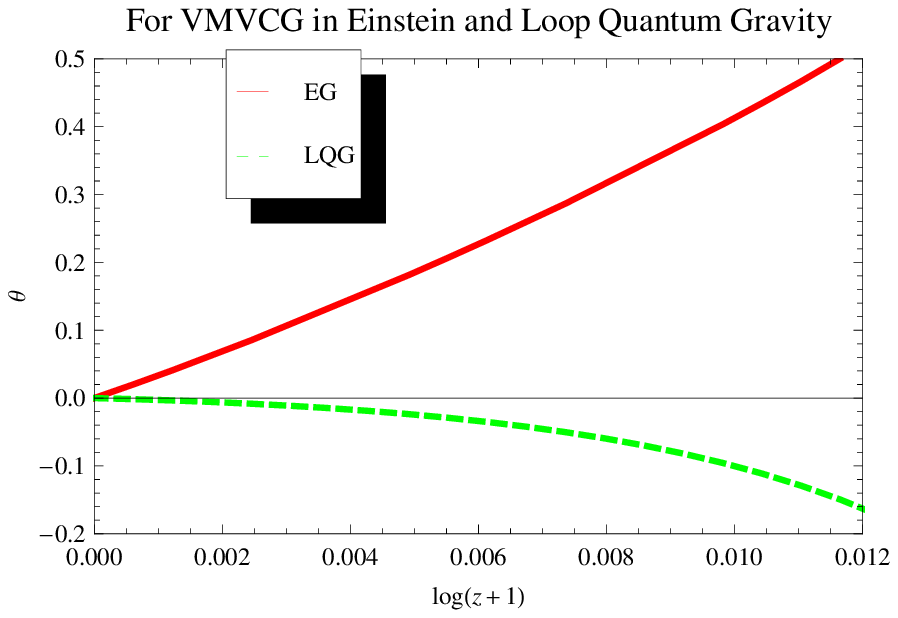}
\caption{Plot of $\theta$ versus $\log(1+z)$.}
\end{minipage}\hspace{3pc}%
\end{figure}
\begin{figure}[h]
\begin{minipage}{14pc}
\includegraphics[width=16pc]{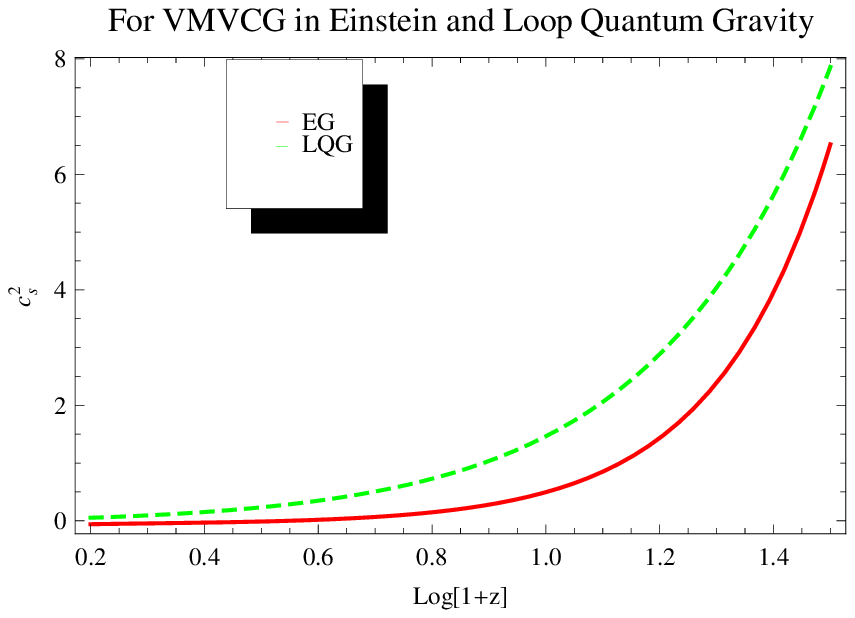}
\caption{Plot of $c^2_s$ versus $\log(1+z)$.}
\end{minipage}\hspace{3pc}%
\begin{minipage}{14pc}
\includegraphics[width=16pc]{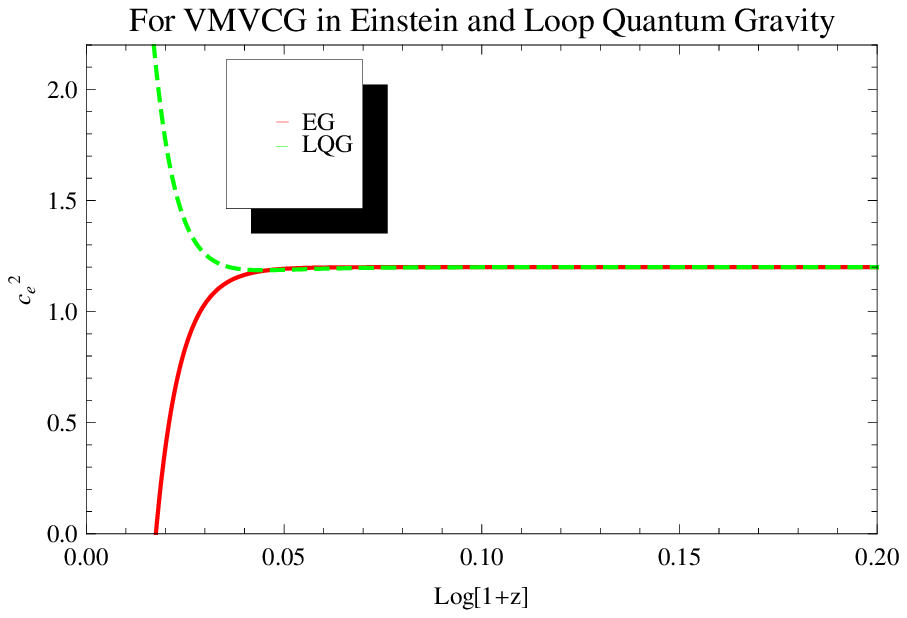}
\caption{\label{label}Plot of $c^2_e$ versus $\log(1+z)$.}
\end{minipage}\hspace{3pc}%
\begin{minipage}{14pc}
\includegraphics[width=16pc]{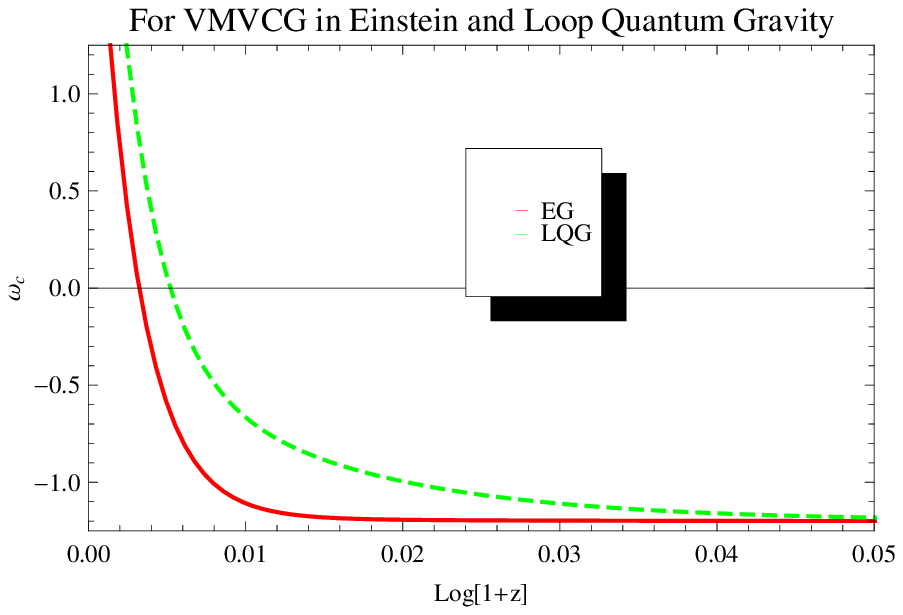}
\caption{\label{label}Plot of $\omega_c$ versus $\log(1+z)$.}
\end{minipage}\hspace{3pc}%
\begin{minipage}{14pc}
\includegraphics[width=16pc]{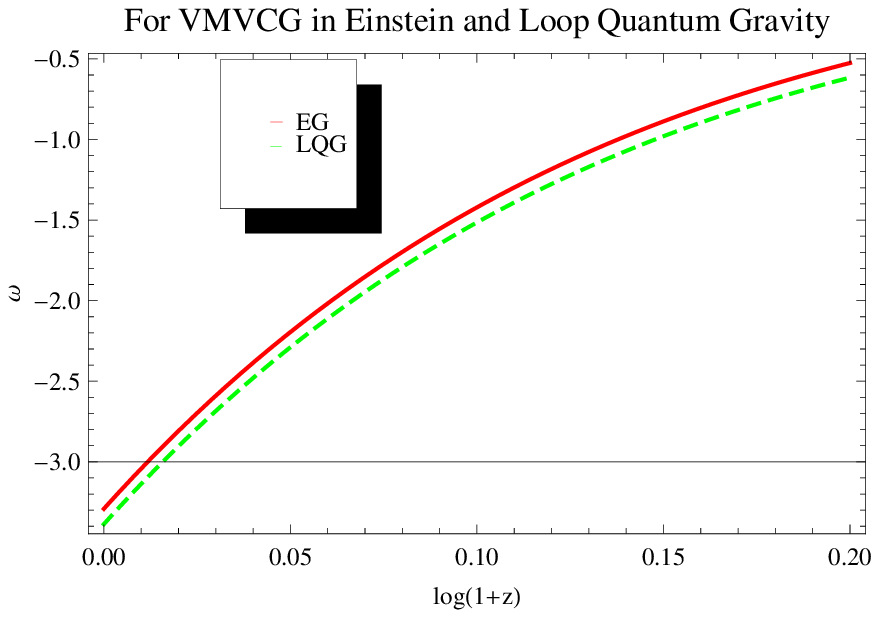}
\caption{\label{label}Plot of $\omega$ versus $\log(1+z)$.}
\end{minipage}\hspace{3pc}%
\end{figure}

We plot the time varying parameters
$\delta,~\theta,~c_{e}^{2},~c_{s}^{2},~\omega_{c}$ and $\omega$
versus $\log(1+z)$ for VMVCG in Einstein gravity and loop quantum
cosmology as shown in Figures \textbf{1-6}. Here, we use
$\Omega=\frac{\rho}{3H^2}$. However, other constants are
$\tilde{A}=1.2,~ B_0=10.2,~
C=0.2,~\xi_{0}=0.01,~\alpha=1.5,~H_{0}=72,~\rho_0=0.23,~\rho_1=2.01$
and $\Omega _{0}=1$. We have chosen the well-known initial condition
as the present value of redshift parameter, i.e., $z=0$ which leads
to $\log[1+z]\rightarrow0$ in our case in all plots. Figure
\textbf{1} shows the evolution of density perturbation and possesses
the increasing behavior with the passage of time. On the other hand,
the other perturbed quantity $\theta$ exhibits the increasing
behavior with the passage of time in Einstein gravity while shows
decreasing behavior in loop quantum cosmology (Figure \textbf{2}).
Figure \textbf{3} indicates that the squared speed of sound remains
positive which leads to the stability of the model at the present as
well as later epoch. In Figure \textbf{4}, the perturbed squared
speed of sound shows increasing/decreasing behavior for
Einstein/Loop quantum cosmology respectively, while approaches to a
positive constant after some interval of time. However, this remains
positive throughout cosmic time which is consistence behavior with
usual squared speed of sound. Figure \textbf{5} (the perturbed EoS)
depicts matter dominated era of the universe at initial epoch while
goes to phantom era at the later epoch by crossing quintessence as
well as $\Lambda$CDM limit. On the other hand, the usual EoS
parameter shows phantom-like universe at the present epoch while
goes towards quintessence region of the universe at the later epoch
(Figure \textbf{6}).

\section{Viscous Generalized Cosmic Chaplygin Gas}

According to earlier studies related to DE corresponding to phantom
era, big rip was the final destination as the time derivative of
scale factor goes to infinity in finite time. By using GCCG model,
big rip singularity can be avoided. The EoS for GCCG has following
form
\begin{equation}\nonumber
p_{d}=-\rho^{-\alpha}\bigg(C+(\rho^{1+\alpha}-C)^{-d}\bigg),\quad
C=\frac{A'}{(1+\omega)}-1
\end{equation}
$A'$ takes either positive or negative constant value, $-l<d<0$ and
$l>1$. The EoS reduces to that of current Chaplygin unified models
for DM and DE in the limit $\omega\rightarrow0$ and satisfies the
conditions:
\begin{itemize}
\item It behaves like de Sitter fluid with $\omega=-1$.
\item It becomes $p=\omega\rho$ as chaplygin
parameter $A' \rightarrow0$.
\item It reduces to the EoS of current chaplygin
unified DM model at high energy density.
\item The evolution of density
perturbations derived from the chosen EoS becomes free from the
pathological behavior of the matter power spectrum for physically
reasonable values of the involved parameter at late time. This EoS
shows a dust era in the past and $\Lambda$ CDM in future.
\end{itemize}
After adding bulk viscosity, above equation becomes
\begin{equation}\nonumber
p_{d}=-\rho^{-\alpha}[C+(\rho^{1+\alpha}-C)^{-d}]-3\xi_{0}H\sqrt{\rho}.
\end{equation}
The solution of conservation equation after substituting above
equation is given by
\begin{eqnarray}\label{11}
\rho=\bigg(C+\bigg(\frac{1}{1-3\sqrt{3}\xi_{0}}+\frac{B}{a^{3(1+\alpha)(1+d)
(1-3\sqrt{3}\xi_{0})}}\bigg)^{\frac{1}{1+d}}\bigg)^{\frac{1}{1+\alpha}}.
\end{eqnarray}
This expression reduces to the following form in terms of redshift,
\begin{eqnarray}\nonumber
\rho=\bigg(C+\bigg(\frac{1}{1-3\sqrt{3}\xi_{0}}+B
(1+z)^{3(1+\alpha)(1+d)(1-3\sqrt{3}\xi_{0})}}\bigg)^{\frac{1}{1+d}}\bigg)^{\frac{1}{1+\alpha}
.\end{eqnarray}
The Hubble parameter turns out to be
\begin{eqnarray}\nonumber
H(z)&=&H_{0}\bigg(\bigg(C+\bigg(\frac{1}{1-3\sqrt{3}\xi_{0}}+B(1+z)^{3(1+\alpha)(1+d)(1
-3\sqrt{3}\xi_{0})}\bigg)^{\frac{1}{1+d}}\bigg)^{\frac{1}{1+\alpha}}\\\label{12}
&\times&\Omega_{0}\bigg)^{\frac{1}{2}}.
\end{eqnarray}
In this case, the EoS parameter becomes
\begin{equation}\label{13}
\omega=-\rho^{-(\alpha+1)}[C+(\rho^{1+\alpha}-C)^{-d}]-\frac{3\xi_{0}H}{\sqrt{\rho}}.
\end{equation}

\subsection{STHC for VGCCG in Einstein Gravity}

For VGCCG, the perturbed EoS parameters takes the following form
\begin{equation}\label{22}
\omega_{c}=\frac{1}{1+\delta}\bigg(-\rho^{-(\alpha+1)}\big(C
+(\rho^{1+\alpha}-C)^{-d}\big)-\frac{3\xi_{0}H}{\sqrt{\rho}}+c^{2}_{e}
\delta\bigg).
\end{equation}
Similarly, the perturbed equation of sound speed turns out to be
\begin{eqnarray}\nonumber
c_{e}^{2}&=&\frac{1}{\rho
\delta}C\big(\rho^{-\alpha}-\rho(1+\delta)^{-\alpha}\big)-
3\xi_{0}H\sqrt{\rho}(1-\sqrt{1+\delta})+\rho^{-\alpha}(\rho^{1+\alpha}-C)^{-d}\\\label{23}
&-&
\rho^{-\alpha}(1+\delta)^{-\alpha}\big(\rho^{1+\alpha}(1+\delta)^{1+\alpha}\big)^{-d}.
\end{eqnarray}
In this scenario, Eqs.(\ref{A20}) and (\ref{21}) leads to
\begin{eqnarray}\nonumber
\frac{d\delta}{dz}&=& \frac{3}{1+z}\bigg(\bigg(\frac{1}{\rho
\delta}C\big(\rho^{-\alpha}-\rho(1+\delta)^{-\alpha}\big)- 3\xi_{0}H
\sqrt{\rho}(1-\sqrt{1+\delta})\\\nonumber
&+&\rho^{-\alpha}(\rho^{1+\alpha}-C)^{-d}-
\rho^{-\alpha}(1+\delta)^{-\alpha}[\rho^{1+\alpha}(1+\delta)^{1+\alpha}]^{-d}\bigg)-\omega\bigg)\delta\\\nonumber
&-& \bigg(1+\omega+\bigg(1+\bigg(\frac{1}{\rho
\delta}C\big(\rho^{-\alpha}-\rho(1+\delta)^{-\alpha}\big)-3\xi_{0}H
\sqrt{\rho}(1-\sqrt{1+\delta})\\\nonumber &+&
\rho^{-\alpha}(\rho^{1+\alpha}-C)^{-d}-
\rho^{-\alpha}(1+\delta)^{-\alpha}\big(\rho^{1+\alpha}(1+\delta)^{1+\alpha}\big)^{-d}\bigg)\bigg)\delta\bigg)
\frac{\theta}{H(z)}\\\nonumber \frac{d\theta}{dz}&=& \frac{\theta
}{1+z}+\frac{\theta^{2}}{3 H(z)}+\frac{3H(z)}{2(1+z)^{2}}\bigg(1+3
\bigg(\frac{1}{\rho
\delta}C\big(\rho^{-\alpha}-\rho(1+\delta)^{-\alpha}\bigg)\\\nonumber
&-& 3\xi_{0}H \sqrt{\rho}(1-\sqrt{1+\delta})
+\rho^{-\alpha}\big(\rho^{1+\alpha}-C\big)^{-d}-
\rho^{-\alpha}(1\\\nonumber &+&\delta)^{-\alpha}
\bigg(\rho^{1+\alpha}(1+\delta)^{1+\alpha}\big)^{-d}\bigg)\bigg)\delta
\Omega.
\end{eqnarray}

\subsection{STHC for VGCCG in Loop Quantum Cosmology}

For VGCCG model in LQC, the Hubble parameter is
\begin{eqnarray}\nonumber
H(z)&=&H_{0}\left[\Omega_{0}(C+(\frac{1}{1-3\sqrt{3}\xi_{0}}+B(1+z)^{3(1+\alpha)(1+d)
(1-3\sqrt{3}\xi_{0})})^{\frac{1}{1+d}})^{\frac{1}{1+\alpha}}\right]^{\frac{1}{2}}\\\nonumber
&\times&\left[1-\left(C+\left(\frac{1}{1-3\sqrt{3}\xi_{0}}
+B(1+z)^{3(1+\alpha)(1+d)(1-3\sqrt{3}\xi_{0})}\right)^{\frac{1}{1+d}}\right)^{\frac{1}{1+\alpha}}\right.\\\label{F2}
&\times&\left.\frac{3\Omega_{0}H_{0}^{2}}{\rho_{1}}\right]^{\frac{1}{2}}.
\end{eqnarray}
\begin{figure}[h]
\begin{minipage}{14pc}
\includegraphics[width=16pc]{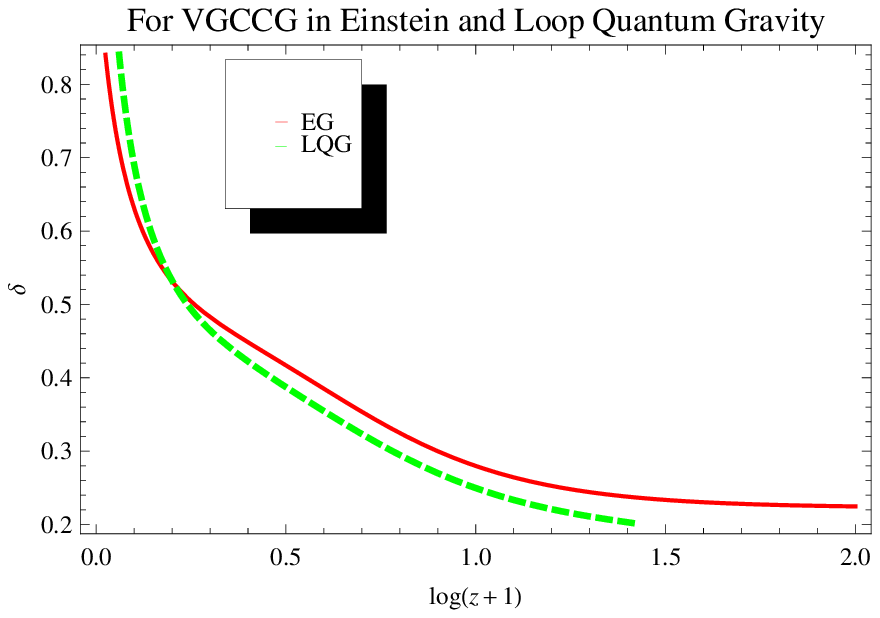}
\caption{\label{label}Plot of $\delta$ versus $\log(1+z)$.}
\end{minipage}\hspace{3pc}%
\begin{minipage}{14pc}
\includegraphics[width=16pc]{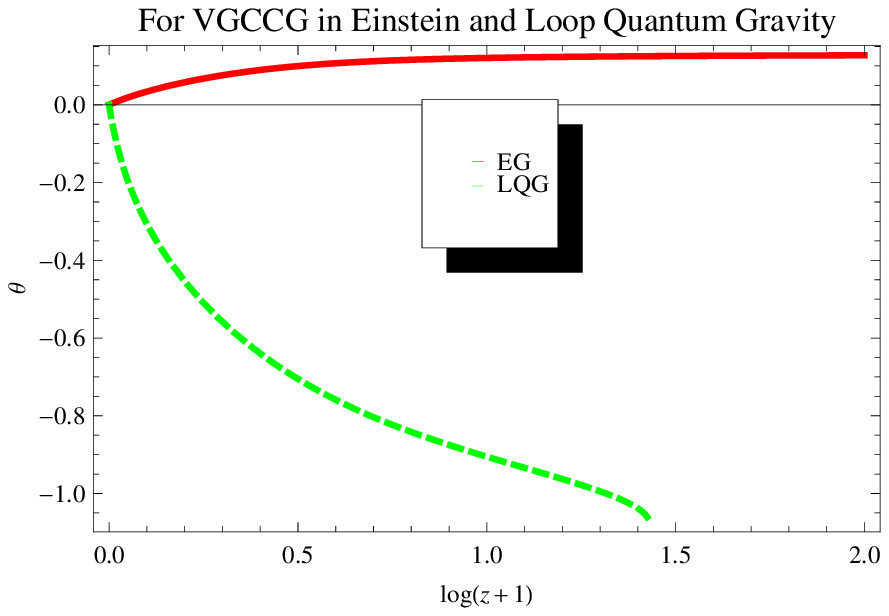}
\caption{Plot of $\theta$ versus $\log(1+z)$.}
\end{minipage}\hspace{3pc}%
\end{figure}
\begin{figure}[h]
\begin{minipage}{14pc}
\includegraphics[width=16pc]{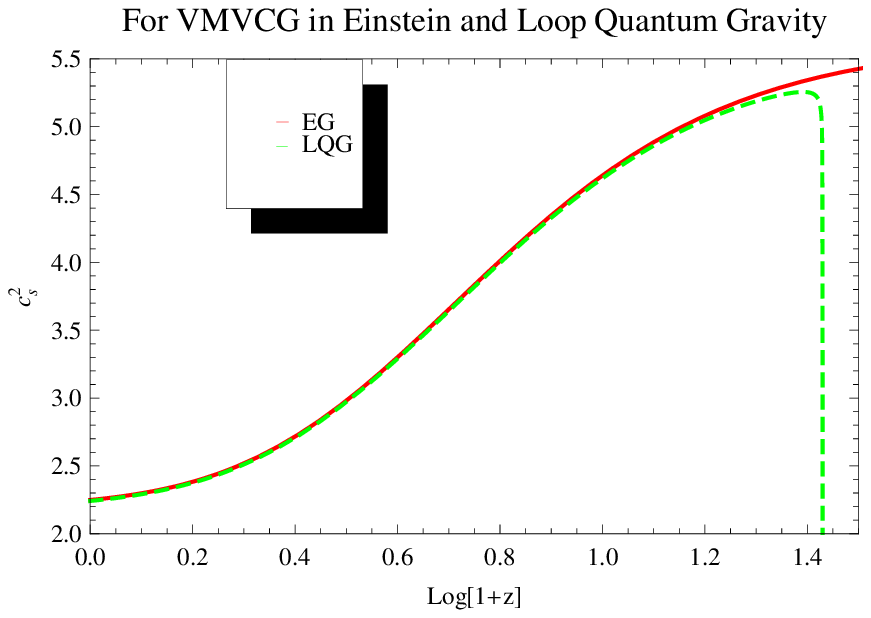}
\caption{Plot of $c^2_s$ versus $\log(1+z)$.}
\end{minipage}\hspace{3pc}%
\begin{minipage}{14pc}
\includegraphics[width=16pc]{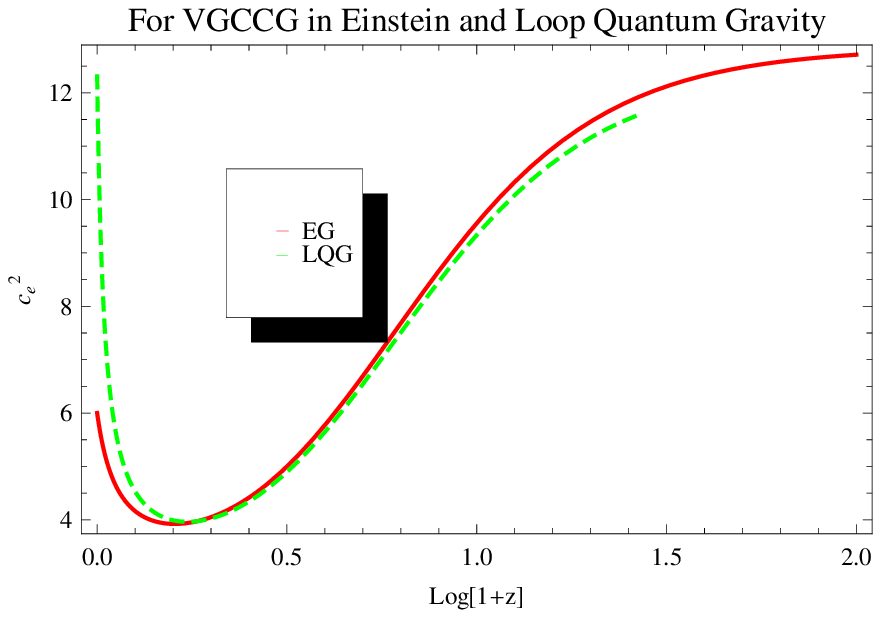}
\caption{\label{label}Plot of $c^2_e$ versus $\log(1+z)$.}
\end{minipage}\hspace{3pc}%
\begin{minipage}{14pc}
\includegraphics[width=16pc]{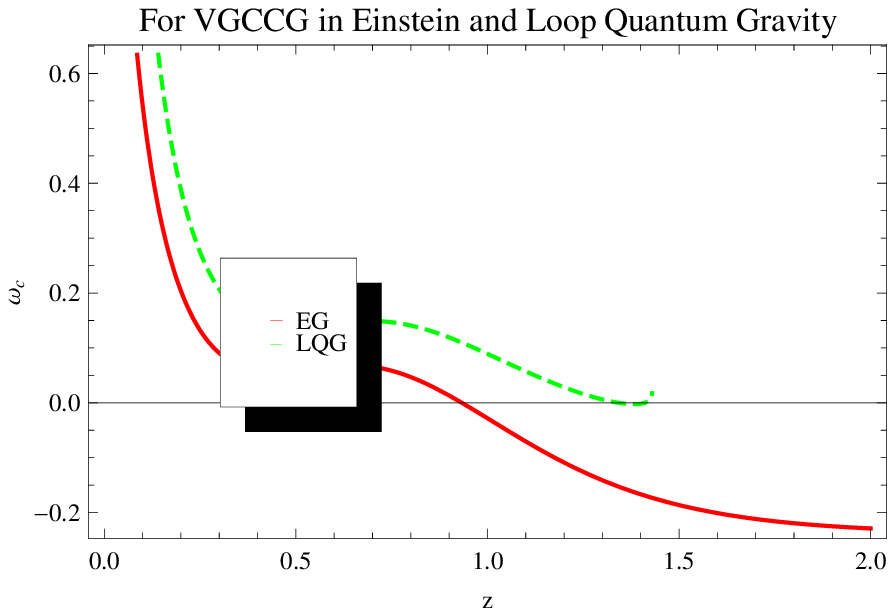}
\caption{\label{label}Plot of $\omega_c$ versus $\log(1+z)$.}
\end{minipage}\hspace{3pc}%
\begin{minipage}{14pc}
\includegraphics[width=16pc]{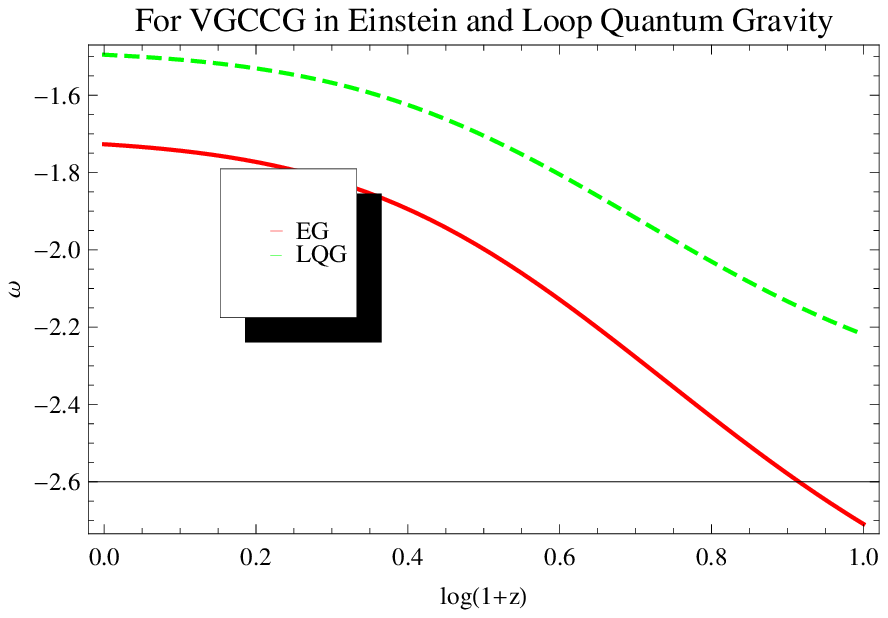}
\caption{\label{label}Plot of $\omega$ versus $\log(1+z)$.}
\end{minipage}\hspace{3pc}%
\end{figure}

For VGCCG in Einstein gravity and loop quantum cosmology, the plots
of the time varying parameters
$\delta,~\theta,~c_{e}^{2},~c_{s}^{2},~\omega_{c}$ and $\omega$
versus $\log(1+z)$ can be obtained by inserting the values of
corresponding expression of $H$ and shown in Figures \textbf{7-12}.
The constant parameters are same as chosen in previous section. It
can be observed from Figure \textbf{7} that the evolution of density
perturbation depicts decreasing behavior with the passage of time.
However, the other perturbed quantity $\theta$ exhibits the same
behavior as observed for VMVCG in Einstein and loop quantum
cosmology (Figure \textbf{8}). We can also see from Figure
\textbf{9} and \textbf{10} that perturbed as well as usual squared
speed of sound remains positive which leads to the stability of the
model at the present as well as later epoch for Einstein/Loop
quantum cosmology. The perturbed EoS (Figure \textbf{11}) depicts
the matter dominated era of the universe at initial as well as later
epoch in Einstein and loop quantum cosmology. The usual EoS
parameter shows the phantom-like universe at the present epoch as
well as at the later epoch (Figure \textbf{12}).

\section{Conclusion}

There exist various cosmological and astrophysical investigations
for the various CG models. These models also used to investigate the
problems of early universe such as inflationary era
\cite{inf1,inf2,inf3}. The accreting phenomenon onto various black
holes has also been discussed with these models. It is found that
the phantom-like behavior of CG models reduces the mass of black
holes. Babichev et al. \cite{acc1} argued that black hole loses its
mass during the accretion of phantom DE onto it. In order to support
this idea, different attempts have been made in the presence of
different phantom-like DE models such as CG, GCG, MCG, VMCG models
etc \cite{acc2,acc3}. Phantom-like behavior of CG is also more
effective in the wormhole physics where the event horizon can be
avoided due to its presence \cite{acc4,acc5}.

Bhar et al. \cite{bh1} investigated the new non-singular model for
anisotropic charged fluid sphere in $(2 + 1)$-dimensional anti
de-Sitter spacetime corresponding to the exterior BTZ spacetime.
They have chosen we choose modified Chaplygin gas in order to solve
the Einstein-Maxwell field equations and show that their model
satisfies all required physical conditions for representing compact
stars. Sharif and Jawad \cite{acc5} have constructed traversable,
asymptotically flat and stable wormhole solutions with the help of
GCCG.

In the present work, we have focused on the top-hat collapse of a
spherically symmetric fluid. We have explored the non-linear
evolution of VMVCG and VGCCG perturbations in classical top-hat
profile by taking the background of flat FRW metric. In the
perturbed region, we have observed the natures of perturbed
quantities like density contrast ($\delta$) and  $\theta$, EoS
parameter ($\omega_{c}$), square speed of sound ($c_{e}^{2}$) for
both chaplygin gas models in Einstein and loop quantum cosmology. We
have also calculated the usual EoS parameter ($\omega$), square
speed of sound ($c_{s}^{2}$) in both scenarios. We have analyzed all
the cosmological parameters graphically versus logarithmic scale of
redshift parameters (Figures \textbf{1-12}). We have summarized our
results as follows
\begin{itemize}
\item \textbf{For VMVCG:}\\ The evolution of
density perturbation exhibited the increasing behavior with the
passage of time for both gravities. However, $\theta$ shows the
increasing behavior with the passage of time in Einstein gravity
while shows decreasing behavior in loop quantum cosmology (Figure
\textbf{2}). The usual as well as perturbed squared speed of sound
have led the stability of VMVCG model in both gravities (Figures
\textbf{3} and \textbf{4}). Figure \textbf{5} (the perturbed EoS)
has suggested the matter dominated era of the universe at initial
epoch while goes to phantom era at the later epoch by crossing
quintessence as well as $\Lambda$CDM limit. The usual EoS parameter
shows phantom-like universe at the present epoch while goes towards
quintessence region of the universe at the later epoch (Figure
\textbf{6}).
\item \textbf{For VGCCG:}\\
It can be observed from Figure \textbf{7} that the evolution of
density perturbation depicts decreasing behavior with the passage of
time. However, the other perturbed quantity $\theta$ exhibits the
same behavior as observed for VMVCG in Einstein and loop quantum
cosmology (Figure \textbf{8}). We have also observed that perturbed
as well as usual squared speed of sound remains positive which leads
to the stability of the model at the present as well as later epoch
for Einstein/LQC (Figures \textbf{9} and \textbf{10}). The perturbed
EoS (Figure \textbf{11}) depicts the matter dominated era of the
universe at initial as well as later epoch in Einstein and loop
quantum cosmology. The usual EoS parameter shows the phantom-like
universe at the present epoch as well as at the later epoch (Figure
\textbf{12}).
\end{itemize}
Finally, it can be concluded that both chaplygin gas models support
the spherical collapse in Einstein as well as loop quantum cosmology
because density contrast remains positive in both cases (Figures
\textbf{1} and \textbf{7}). Also, the perturbed EoS remains positive
at the present epoch as well as near future (Figures \textbf{5} and
\textbf{11}) in both cases. It can be also remarked that some of our
results are consistence with \cite{9,10,top1,17,17aa,13}.

\section*{Appendix}

For the perturbed region, the basic equations are
\begin{eqnarray}\label{A17}
\dot{\rho_{c}}= -3h(\rho_{c}+p_{c}),\quad\frac{\ddot{r}}{r}=
\frac{-4\pi G(\rho_{c}+3 p_{c})}{3},
\end{eqnarray}
where $r$ is the local scale factor and $h=\frac{\dot{r}}{r}$
relates to local expansion rate in the STHC framework
\begin{equation}\label{A18}
h= H + \frac{\theta}{3a},
\end{equation}
where $\theta=\nabla.v$ is the divergence of the peculiar velocity
$v$. The dynamical equations of density contrast $\delta$ and
$\theta$ can be calculated as \cite{top1}
\begin{eqnarray}\label{A19}
\delta'&=& -\frac{3}{a}(c_{e}^{2}-\omega)\delta-
[1+\omega+(1+c_{e}^{2})\delta]\frac{\theta}{a^{2}H },\\\label{19}
\theta'&=& -\frac{\theta
}{a}-\frac{\theta^{2}}{3a^{2}H}-\frac{3H}{2}(1+c_{e}^{2})\delta
\Omega,
\end{eqnarray}
where $\Omega =\frac{\rho}{3H^{2}}$ and prime represents the
derivative with respect to scale factor $a$. The above equations can
be rewritten in terms of $z$ as
\begin{eqnarray}\label{A20}
\frac{d\delta}{dz}&=& \frac{3}{1+z}(c_{e}^{2}-\omega)\delta-
\big(1+\omega+(1+c_{e}^{2})\delta\big)\frac{\theta}{H(z)}\\\label{21}
\frac{d\theta}{dz}&=& \frac{\theta }{1+z}+\frac{\theta^{2}}{3
H(z)}+\frac{3H(z)}{2(1+z)^{2}}(1+3 c_{e}^{2})\delta \Omega.
\end{eqnarray}


\begin{thebibliography}{43}
\bibitem{1} Riess, A.G., et al.: Astron. J. \textbf{116}(1998)1009.
\bibitem{2} Perlmutter, S., et al.: Astrophys. J. \textbf{517}(1999)565.
\bibitem{3} Sahni, V. and Starobinsky, A.: Int.J. Mod. Phys. D \textbf{373}(2000)9.
\bibitem{4} Caldwell, R.R.,  Dave, R. and  Steinhardt, P.J.: Phys. Rev. Lett. J. \textbf{80}(1998)1852.
\bibitem{5} Caldwell, R. R.:  Phys. Lett. B \textbf{545}(2002)23.
\bibitem{6} Chaplygin, S.: Sci. Mem. Moscow Univ. Math. Phys.
\textbf{21}(1901)1.
\bibitem{7} Kowalski, M., et al.: Astrophys. J. \textbf{686}(2008)749.
\bibitem{8} Tsujikawa, S.: arXiv:1004.1493.
\bibitem{9} Li, W. and  Xu, L.: Eur. Phys. J. C \textbf{73}(2013)2471.
\bibitem{10} Li, W. and  Xu, L.: Eur. Phys. J. C \textbf{74}(2014)2765.

\bibitem{18} Gonzalez-Diaz, P.F.: Phys. Rev. D \textbf{68}(2003)021303(R).


\bibitem{11} Gunn, J.E. and Gott, J.R.: Astrophys. J. \textbf{176}(1972)1
\bibitem{12} Padmanabhan, T. \emph{Structure formation in the
universe} (Cambridge Univ. Press, 1993).

\bibitem{top1} Fernandes, R.A.A., Carvalho, J.P.M.D., Kamenshchik, A.Y., Moschella, U. and Silva, A.D.: Phys. Rev.
D \textbf{89}(2014)083533.


\bibitem{top2} Bilic, N.,et al.: JCAP \textbf{11}(2004)8.

\bibitem{top3} Maccio, A. V. et al.: Phys.
Rev. D \textbf{69}(2004)123516; Aghanim, N., et al.: A\&A
\textbf{496}(2009); Baldi, M., et al.: MNRAS \textbf{403}(2010),
1684; Li, B., et al.: Astrophys. J. \textbf{728}(2011)109.



\bibitem{14} Fabris, J.C., Gonclaves, S.V.B. and de Souza, P.E.:  Gen. Relav. Grav. \textbf{34}(2002)53.
\bibitem{15} Carturan, D. and  Finelli, F.: Phys. Rev. D \textbf{68}(2003)103501.

\bibitem{17} Carames, T.R.P., Faris, J.C. and Velten, H.E.S.:  Phys. Rev. D
\textbf{89}(2014)083533.

\bibitem{17aa} Karbasi, S. and Razmi, H.: Int. J. Mod. Phys. D
\textbf{24}(2015)1550050.

\bibitem{13} Debnath, U. and Jamil, M.:  arXiv:1501.00486.


\bibitem{19} Kamenshchik, A., Moschella, U., Pasquier, V.: Phys. Lett. B \textbf{511}(2001),
265.
\bibitem{20} Bento, M.C., Bertolami, O. and Sen, A.A.: Phys. Rev. D
\textbf{66}(2002)043507.
\bibitem{21} Gorini, V., Kamenshchik, A. and Moschella, U.: Phys. Rev. D
\textbf{67}(2003), 063509.
\bibitem{22} Rodrigues, D.C.: Phys. Rev. D \textbf{78}(2008)063013.
\bibitem{23} Jamil, M. and Rashid, M.A.: Eur. Phys. J. C
\textbf{60}(2009)141; DePaolis, F., Jamil, M. and Qadir, A.: Int. J.
Theor. Phys. \textbf{49}(2010)621.
\bibitem{24a} Eckart, C.: Phys. Rev. \textbf{58}(1940)919.
\bibitem{25bb} Landau, L.D., Lifshitz, E.M.: \emph{Fluid Mechanics}
(Butterworth Heinemann, Oxford, 1987).
\bibitem{26} Xu, Y.D., et al.: Astrophys Space Sci \textbf{339}(2012)31.


\bibitem{36} M. Bojowald, Living Rev. Relativity 11 (2008), 4; K. Xiao,
Xiao-Kai He, Jian-Yang Zhu, Phys Lett B 727, 349 (2013); X-M Zhang,
J-Y Zhu, Phys. Rev. D. 87, 043522 (2013); A. Barrau, L. Linsefors,
JCAP 12 (2014) 037.

\bibitem{37} M. Jamil, D. Momeni, M.A. Rashid, Eur. Phys. J. C 71, 1711
(2011); K. Karami, M. Jamil, N. Sahraei, Phys. Scr. 82 (2010)
045901; H.M. Sadjadi, M. Jamil, Gen. Rel. Grav. 43, 1759 (2011).

\bibitem{38} M. Bojowald, G. M. Hossain, Phys. Rev. D 77, 023508 (2008); J.
Mielczarek, JCAP 0811, 011 (2008); J. Grain, T. Cailleteau, A.
Barrau, A. Gorecki, Phys. Rev. D 81, 024040 (2010)

\bibitem{39} T. Cailleteau, A. Barrau, Phys. Rev. D 85, 123534 (2012); Yu
Li, J-Y Zhu, Phys. Rev. D 85, 023515 (2012); M. Bojowald, G.
Calcagni, S. Tsujikawa, Phys. Rev. Lett. 107, 211302 (2011).

\bibitem{L22} Bojowald, M.: Living Rev. Rel. \textbf{8}(2005)11.

\bibitem{L23} Ashtekar, A., Bojowald, M. and
Lewandowski, J.: Adv. Theor. Math. Phys. \textbf{7}(2003)233.

\bibitem{L24} Ashtekar, A.: AIP Conf. Proc. \textbf{861}(2006)3.

\bibitem{L25} Rovelli, C.: Living Rev. Rel. \textbf{1}(1998)1.

\bibitem{L26} Ashtekar, A. and Lewandowski, J.: Class. Quant. Grav. \textbf{21}, R53
(2004).

\bibitem{L27} Rovelli, C.: Quantum Gravity, Cambridge University Press,
Cambridge (2004).

\bibitem{L27a} Wu, P. and Zhang, S. N., 2008, JCAP 06, 007.

\bibitem{L27b} Chen, S., Wang, B. and Jing, J., 2008, Phys. Rev. D 78, 123503.

\bibitem{L29} Jamil, M. and Debnath, U., 2011, Astrophys Space Sci. 333, 3. [27]

\bibitem{L30} Fu, X., Yu, H. and Wu, P., 2008, Phys. Rev. D 78, 063001.

\bibitem{L31} Chakraborty, S., Debnath, U. and Ranjit, C.: Eur. Phys. J. C
\textbf{72}(2012)2101.


\bibitem{inf1} del Campo, S. and Herrera, R.: Phys. Lett. B\textbf{665}(2008)100.

\bibitem{inf2} Herrera, R., Olivares, M. and Videla, N.: Eur. Phys. J. C
\textbf{73}(2013)1.

\bibitem{inf3} Setare, M.R. and Kamali, V.: Phys. Rev. D \textbf{91}(2015)123517.


\bibitem{acc1} Babichev, E., Dokuchaev, V. and Eroshenko, Y.: Phys. Rev. Lett.
\textbf{93}(2004)021102.

\bibitem{acc2} Jawad, A. and Shahzad, M. U.: Eur. Phys. J. C  \textbf{76},
123 (2016); Babichev, E. et al.: Phys. Rev. D
\textbf{78}(2008)104027; Jamil, M.: Eur. Phys. J. C
\textbf{62}(2009)325.

\bibitem{acc3} Bhadra, J. and Debnath, U.:
Eur. Phys. J. C \textbf{72}(2012)1912.

\bibitem{acc4} Sharif, M. and Jawad, A.: Eur. Phys. J. Plus \textbf{129}, 15
(2014); Lobo, F.S.N.: Phys. Rev. D \textbf{71}(2005)124022; Lobo,
F.S.N.: Phys. Rev. D \textbf{71}(2005)084011; Sushkov, S.: Phys.
Rev. D \textbf{71}(2005)043520.

\bibitem{acc5} Sharif, M. and Jawad, A.: Eur. Phys. J. Plus \textbf{129}(2014)15.


%
%


\bibitem{bh1} Bhar, P et al.: Astrophys Space Sci \textbf{360}(2015)32.


\end{thebibliography}
\end{document}